\documentclass[twocolumn,showkeys,preprintnumbers,amsmath,amssymb]{revtex4}

\usepackage{graphicx}
\usepackage{natbib}
\usepackage[T2A]{fontenc}
\usepackage{dcolumn}
\usepackage{bm}
\usepackage{hyperref}
\usepackage{float}
\restylefloat{table}
\newcommand{\RNumb}[1]{\uppercase\expandafter{\romannumeral #1\relax}}

\begin{document}
	
	\title{Measurements of thermal relaxation of the OGRAN underground setup. }

	\author{$^{1,2}$Gavrilyuk Y.M.}

	\author{$^1$Gusev A.V}%
		\author{$^3$Kvashnin N.L.}%
			\author{$^3$Lugovoy A.A.}%
				\author{$^1$Oreshkin S.I.}%
					\author{$^1$Popov S.M.}%
	\author{$^1$Rudenko N.V.}%
	\author{$^1$Semenov V.V.}%
		\author{$^1$Syrovatsky I.A. }%

	\affiliation{%
		$^1$Sternberg Astronomical Institute, Moscow State University, Moscow, 119991 Russia 
	}%
	\affiliation{%
	$^2$Institute for Nuclear Research RAS, Baksan Neutrino Observatory
}%
	\affiliation{%
	$^3$Institute of Laser Physics SB RAS
}%

	\begin{abstract}
		\mbox{Abstract} -- An upgraded version of the OGRAN - combined optical-acoustic gravitational wave detector- has been investigated in a long-term operation mode. This installation, located at the Baksan Neutrino Observatory (BNO) INR RAS, is designed to work under the program for detecting collapsing stars in parallel with the neutrino detector: - Baksan Underground Scintillation Telescope (BUST). Such joint search corresponds to the modern trend for a development of ``multi-messenger astronomy''. In this work the effects of thermal relaxation OGRAN are experimentally investigated using passive and active thermal stabilization systems in the underground laboratory BNO PK-14.
	\end{abstract}
	\maketitle	

\section[Introduction ]{Introduction}
	The modern trend of ``multi-channel'' astronomy  or ``multi-messenger astronomy'' (MMA)  implies the parallel observation of extraterrestrial relativistic objects through radiation channels of various physical nature: wide frequency range electromagnetic radiation, radiation in the form of cosmic ray of elementary particles, including neutrino fluxes and, finally, the recently opened channel of gravitational-wave radiation. It should be emphasized that due to the weak scattering of gravitational waves (GW) and neutrinos, the MMA makes it possible to study astrophysical phenomena in vary distant space. The first source from which both gravitational and electromagnetic radiation (in X-ray, gamma, optical and infrared ranges) had been recorded was the merger of two neutron stars at a distance of about 40 Mpc from the Earth (event GW170817). It was recorded by the LIGO and VIRGO gravitational detectors and, at the same time, by the Fermi and Integral satellites registering gamma-ray bursts \cite{linkk1}. After 10 hours, optical telescopes (including the MASTER network of telescopes \cite{linkk2}) detected a supernova explosion in the NGC galaxy in the localization zone of the gamma - gravitational source.
	
	To date, all GW bursts recorded have been associated with the processes of merging of binary relativistic objects which generate so-called ``chirp signals'', the shape of which is predicted theoretically. GW radiation accompanying the collapse of massive stars does not possess such structural universality. This complicates the possibility of detecting GW radiation from collapsars. This is also the reason why their gravitational signals have not yet been detected. 
	
	The MCA strategy can increase the probability of such registration, taking into account that the collapse process should be accompanied by neutrino radiation  \cite{linkk3}. In this regard the joint program of MSU and RAS institutes was formulated to detect neutrino and gravitational radiation from objects in our Milky Way Galaxy using the Baksan Underground Scintillation Telescope (BUST) and the opto-acoustic gravitational antenna (OGRAN) located in the underground laboratory of the Baksan Neutrino Observatory of INR RAS.
	
	The OGRAN gravitational antenna uses a combination of solid-state resonance acoustic bar and optical interferometric principles for detecting GWs \cite{linkk4}\cite{linkk5}. The central element of the antenna is a massive cylindrical bar (M=2000 kg, L=2.0 m) made of an aluminum alloy with a central axial tunnel. The plane and spherical mirrors are attached to the ends of the bar forming the Fabry-Perot cavity. This design introduces two new qualitative factors:
	1) the impact of GW on two degrees of freedom, acoustic and optical that creates a complex response structure, facilitating its filtering;
	2) a low-noise optical registration system makes it possible to achieve without cooling a sensitivity $\sim10^{-19}$ (in terms of space metric variation) typical for cryogenic antennas. The sensitivity, however, remains limited at the level of $10^{-20}$ by the thermal acoustic noise of the detector outside of its resonant region. In 2020, a significant upgrade of the OGRAN antenna was carried out, which made it possible to bring this setup into the mode of long-term continuous measurements \cite{linkk6}.
	
	The BUST neutrino scintillation telescope \cite{linkk7} is located at an effective depth of 850 m in the units of water equivalent. The setup consists of 3184 scintillation counters, the total mass of the scintillator is 330 tonne. The main reaction for detecting neutrinos is the process of inverse beta decay: $\bar{v}_e + p \rightarrow n + e^+$. At an average antineutrino energy $E_\nu \sim 10$ MeV, the positron energy usually is enclosed in the volume of one counter. In this case, the registration signal looks like a series of events from single triggering of counters (the threshold energy of the counter is 8 MeV). During the entire observation period on the BUST under the ``search for collapsars'' program (27 years), no one event exceeding the stochastic counting background was registered. The experimental estimate of the probability of the gravitational collapse in the Galaxy (at a distance of 10 kpc) or the random "rate of events" only on the base of the neutrino channel currently is set as 0.074 per year.
	
	With the addition of the gravitational-wave channel data, further refinement of this estimate is awaited up to the theoretically expected value of $\sim$ 0.01 year$^{-1}$ (the random frequency of the supernova emergence in a single galaxy). An associated task for the planned two-channel search for collapsars is the elaboration of a statistical algorithm for joint processing of data taken from BUST and OGRAN \cite{linkk8}, based on the temporal correlation of GW and neutrino signals predicted by different collapse models \cite{linkk9},\cite{linkk10},\cite{linkk11}. 
	In this paper we present the results of long-term experimental studies of OGRAN thermal relaxation in the regime of passive and active thermal stabilization.
	
\section[Continuous observation mode]{Continuous observation mode }	
	The current technical state of the OGRAN antenna allows continuous long-term observations of the gravitational gradient background in the vicinity of the resonance frequency at 1.3 kHz. The developed software allows remote control of the whole setup and its main parameters are actively monitored. Optical modes of the FP resonators can be automatically found and maintained. Remote access to the main computer of the setup is carried out via the Internet using additional local networks BNO.
	
	At the moment, there have been no serious restrictions on the bandwidth of the control channel. In full-screen mode, with the output of all signals in the form of dynamically updated graphs and images of optical modes in both channels (FP - bar and FP - discriminator), a bandwidth of 5-10 Mbit/s is sufficient. For a remote operator, the efficiency of setup control depends on the propagation time of the communication signals (parameter Round Trip Time (RTT)). When the control centers are located in Moscow or Novosibirsk, the RTT estimate is $\sim$50-100 ms, which satisfies the conditions for comfortable control (in the absence of the need to quickly change the operating modes of the setup). An important function of the automatic control system is its ability to restore the operating mode after its unexpected unlock as a result of intense external disturbances. The software for this function was developed earlier \cite{linkk8}. However, for a full-operated automatic mode, it is necessary to make additional changes to the OGRAN photo-detection (signal registration) system. At the moment, recovery after failures is carried out manually by the operator.
	
	The OGRAN antenna was launched into the continuous observation mode on March 1, 2021. As noted above, this was preceded by an upgrade, including the replacement of photodetectors in both channels, the installation of combined devices consisting of photodetectors for registering the intensity of the transmitted radiation and video cameras for observing the shape of the optical mode. These elements are created for automatically search of the optimal working point position and maintain the setup in the continuous operation mode. 
	The Fig. 1 shows an image of a part of the window (formed by the control program) with the inputs from the video cameras and the lock system status in both channels.

	\begin{figure}[h!]
		\centering
		\includegraphics[width=1\linewidth]{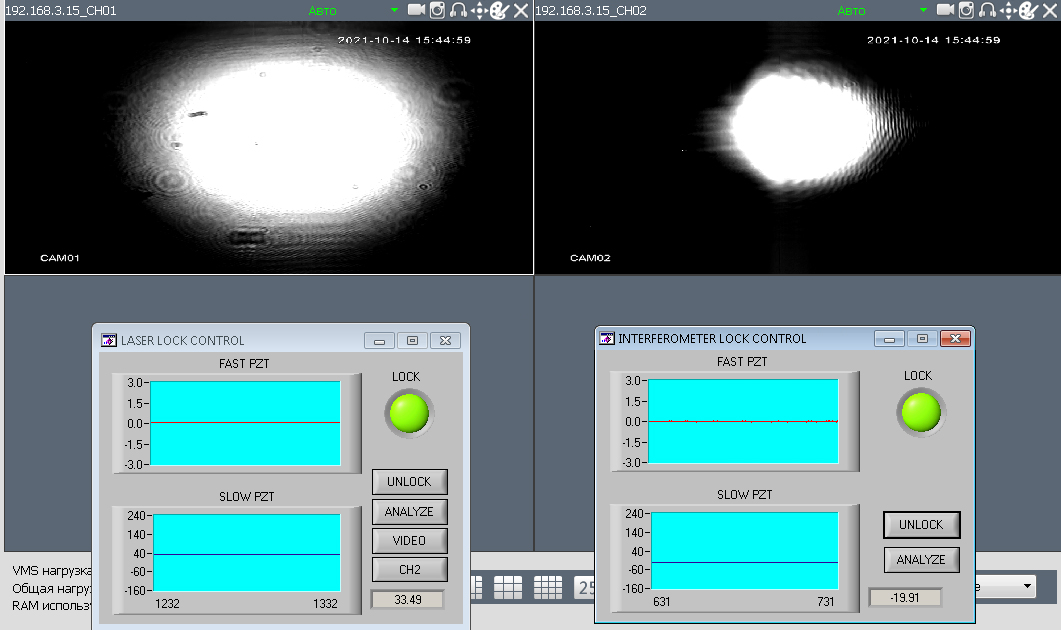}
		\caption{View of the control panel window formed by the control program of the OGRAN installation in the full capture regimes (main optical modes are locked in both channels).}
		\label{fig:pic1}
		\centering
	\end{figure}

\section[Problems of the continuous long observation ]{Problems of the continuous long observation  }	
During observations, important aspects were revealed that affect the long-term confident retention of the full capture operating mode in the system. The main disturbing factor is variations in the temperature background of the laboratory room despite the installed controlled heating system, which includes proportional-integral-differential controller (PID) with 0.05 degrees adjustment accuracy. The temperatures in three zones are independently controlled (at the entrance to the underground laboratory room, in the laboratory itself and directly under the chamber of the gravitational detector). Nevertheless, it was turned out that even thermal radiation from the operator's body has a significant influence. After the visit to the laboratory by the service personnel a stable temperature background had being established for several days. 
It is obvious that defining of the characteristic thermal relaxation times of the setup as a whole and of its individual parts is important to ensure the regime of long-term observations.

Structurally, the OGRAN antenna consists of two parts: the detector arm and the discriminator arm. Both contain Fabry-Perot (FP) interferometers and are combined by a common laser pumping source (see the diagram in \cite{linkk4},\cite{linkk5}). Thermal variations in the length of a solid-state detector change the optical eigen frequency of the FP cavity installed on it. Feedback systems track these changes and control the wavelength of the laser radiation so that the laser pump frequency always matches the FP eigen frequency. The laser frequency is changed by applying a control voltage to the piezoceramic package, on which one of the mirrors of the laser resonator is fixed. Thus, the value of this voltage is proportional to the length of the solid-state detector, i.e. gives information about its thermal state. In the second arm, the FP resonator is mounted on a smaller cylindrical bar made of sitall, a material with low thermal conductivity and thermal expansion coefficients. The pump frequency variations caused by variations in the FP cavity length of the first channel, as well as the failure of the optical FP resonance, are here compensated by the feedback voltage applied to the piezoelectric package, on which one of the FP mirrors is fixed. Thus, this voltage is proportional to the effective optical length of the discriminator.

To keep the operating point at the top of the optical resonance, the Pound - Drever - Hall technique \cite{linkk12} is used in both arms. The needed laser frequency modulation realized by using the single extracavity EOM at the frequency of 10.7 MHz, where the technical laser noise is already lower than the Poisson noise at operating powers. By measuring the voltage on the piezoelectric transducer (PZT) controlling the laser radiation frequency and knowing the distance between the sideband components of the modulated pump ($2\cdot10.7 MHz = 21.4 MHz$), it is possible to determine the conversion factor of the control voltage to the laser frequency. The tuning range in the first channel (bar arm) is determined by the range of possible voltages at the high-voltage amplifier (HVA) driving the PZT inside the laser and is approximately 430 volts. For the first channel's Fabry-Perot resonator (bar solid-state detector with L=2m), the free spectral range is FSR=c/2L=75 MHz. In the laser tuning range, there are 23 TEM$_{00}$ optical modes, i.e. the full tuning range is 1725 MHz.

In the second channel, the feedback circuit adjusts the length of the reference FP cavity to the laser frequency using piezoelectric ceramics under one of the cavity mirrors. For a second channel's cavity length equal to 45 cm, the distance between TEM$_{00}$ modes is FSR=167 MHz. In the entire tuning range of the second channel's HVA (430 volts), only two TEM$_{00}$ modes are present and the interval between them corresponds to the voltage of 200 V. Therefore, the total tuning frequency range of modes in the second channel is 335 MHz.
To understand how two Fabry-Perot resonators will behave (one on an aluminum bar with a length of 2 m, and the other on a sitall one with a length of 45 cm), it is necessary to compare the thermal diffusivity coefficients for these two materials, on which a process of heat propagation inside the resonators depends on. The external temperature change for both resonators is supposed the same.

Thermal diffusivity ($m^2\cdot s^{-1}$) is a physical quantity that characterizes the rate of the heat conduction of the medium in no equilibrium thermal processes. Numerically, it is equal to the ratio of thermal conductivity ($W\cdot m^{-1}\cdot K^{-1}$) to specific heat capacity at constant pressure ($J\cdot kg^{-1}\cdot K^{-1})$ and the density of material ($kg\cdot m^{-3}$).

$$\alpha = \frac{\lambda}{c\cdot\rho}$$

Calculations show that the thermal diffusivity for sitall is 160 times lower than for aluminum. In this case, it should be expected that with a constant temperature background in the surrounding room, the main acoustic detector (bar) relaxes (stabilizes) in temperature much faster than the sitall discriminator. Fig. 2 shows a graph of the dependence of the voltages on piezoelectric ceramics in both channels from time to time (two weeks after starting the setup in the continuous observation).

	\begin{figure}[h!]
	\centering
	\includegraphics[width=1\linewidth]{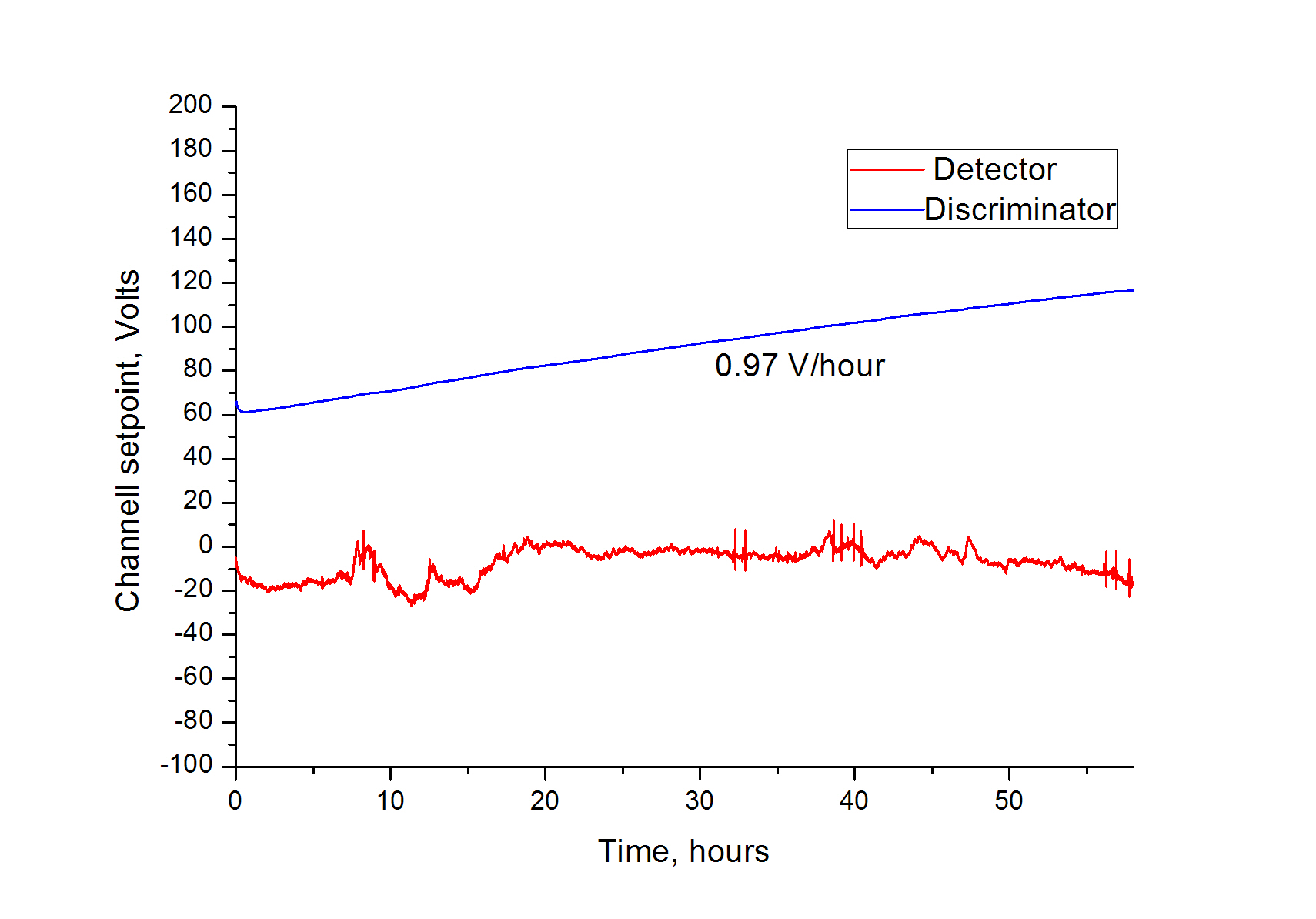}
	\caption{Evolution of control voltages in both channels of the OGRAN setup during two days (and two weeks later after finishing the service mode)
	}
	\label{fig:pic2}
	\centering
\end{figure}
From the figure, we can conclude that with the first channel already sufficiently stabilized (red curve), the second channel (blue curve) continues to drift slowly in one direction (in this case, corresponding to an increase in temperature and, consequently, to a lengthening of the resonator). The splashes on the red curve correspond to the passage of the service electric locomotive through the tunnel. In addition, the level of technical noise is well distinguished between daytime and nighttime. 

For a qualitative check of the time to establish a stable temperature regime on a setup operating in full capture, an experiment was carried out to study the effect of variations in the total temperature background on the stability of capture retention in both channels. In one day the controlled background heating (temperature stabilization) was turned off, and the temperature background began to slowly decrease (over a period of several days) from 25 degrees Celsius to about 22 degrees - the temperature typical of OGRAN laboratory room at a depth of 1350 m. from the entrance into the tunnel this time of year. 

The experiment showed that such a change (by 3 degrees) in the temperature background is fatal for the full capture regime in both channels. It turned out that the position of the operating point pass through the full dynamic range of the HVA many times. Each time at the end of this range, the control of the operation point breaks down. Full capture mode has to be re-established by operator. Fig. 3 shows dependences of the HVA output voltages on time (in hours) from the moment when the temperature stabilization was turned off.

	\begin{figure}[h!]
	\centering
	\includegraphics[width=1\linewidth]{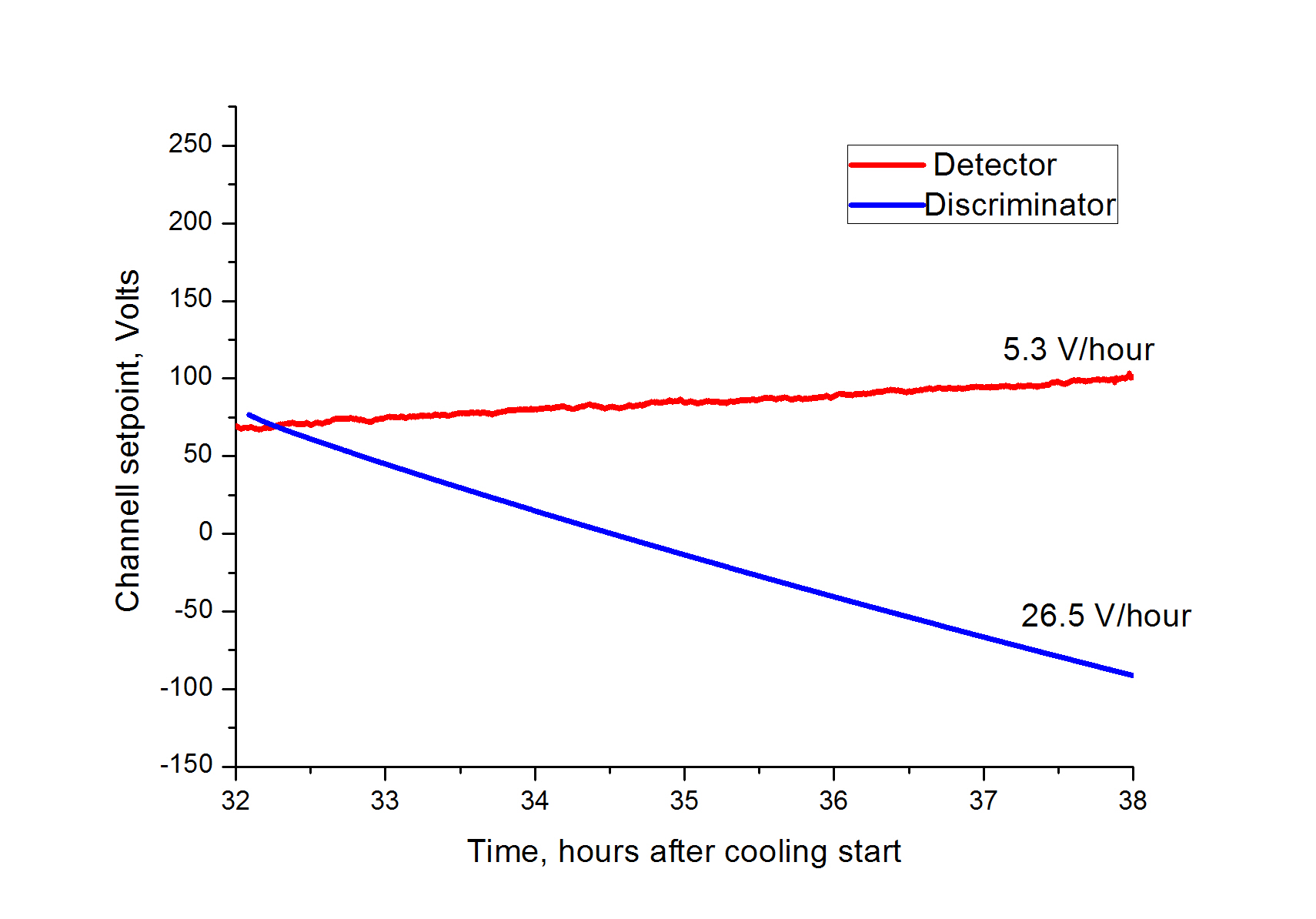}
	\caption{Dependence of the control voltages in both channels of the OGRAN setup on the time on the second day after the temperature stabilization is turned off (both channels are floating, but the capture is maintained)
	}
	\label{fig:pic3}
	\centering
\end{figure}
The solution to the problem of establishing the temperature in the body of both detectors is reduced to solving the partial differential equation:

$$\frac{\partial T}{\partial t} = \alpha \frac{\partial^2 T}{\partial x^2}$$
Here $T$ is the temperature, $\alpha$ is the thermal diffusivity. Depending on the choice of the initial and boundary conditions, the solutions are different (see, for example, \cite{linkk13}). However, the temperature settling time can be qualitatively estimated from the exponential argument in the final solution for a rod with an isolated lateral surface and a given heat transfer to the ends. In the simplest model, one can take the length of such rod equal to the distance from the lateral surface of the detector to the surface of the inner channel. The characteristic time of temperature established in such model

$$\Delta \tau \cong \frac{l^2}{\pi^2\alpha}$$
Here $l$ is the distance from the lateral surface of the detector to the central channel surface, $\alpha$ is the thermal diffusivity. Substituting the values for the large detector $l=30 cm$, $\alpha=8.4\times10^{-5} m^2/s$, we obtain the temperature settling time of the order of two minutes. For a discriminator made of sitall with $l=10 cm, \quad \alpha=5.3\times10^{-7} m^2/s$, the temperature settling time is about 30 minutes. Heat exchange is carried out mainly due to radiation and thermal conductivity of residual gases. Calculation methods can be found, for example, in \cite{linkk14}. At a sufficiently high vacuum ($\sim10^{-5} \ torr)$, when the free path of the residual gas molecules significantly exceeds the distance from the chamber wall to the outer surface of the detector, the heat exchange of residual gases becomes negligible in comparison with the heat transfer by radiation. The radiation flux between two surfaces with temperatures $T$ and $T_0$ is defined as:
$$\frac{dQ}{dt} = -\varepsilon_n\sigma A_1(T^4 - T^4_0)$$

Here $dQ = c\cdot m\cdot dT$ is the change in the internal energy when the temperature changes by $dT$, c is the heat capacity of the detector material, $m$ is the detector mass, $\sigma$ is the Stefan-Boltzmann constant, $A_1$ is the detector surface area, $\varepsilon_n$ is the emissivity (measure of black body approximation). Solving the ordinary differential equation of the first order with the boundary condition $T=T_0$ at $t=0$, we obtain the temperature relaxation time

$$\tau = \frac{cm\left( \frac{1}{T^3} - \frac{1}{T^3_0}\right) }{3\varepsilon_n\sigma A_1}$$

The change in temperature from 300 K to 301 K of the main detector with a mass of $m = 2000 kg$ with a side surface area of $8.8 m^2$ is about 13 hours, if we take the emissivity equal to 0.01. It was experimentally found that the time of complete thermal relaxation of the setup from the moment when the background heating was switched on until a stable temperature regime reached is about 700 hours (approximately one month). Such times are apparently due to much slower relaxation in the reference cavity (discriminator).

The stability of keeping the operating point in the full capture mode can also be affected by the change in the refractive index due to the changes of the vacuum. For experimental verification of this hypothesis, we carried out an experiment on pumping out the reference cavity chamber with simultaneous observation of the signals of holding the operating points. The results are shown in Fig. 4.
	\begin{figure}[h!]
	\centering
	\includegraphics[width=1\linewidth]{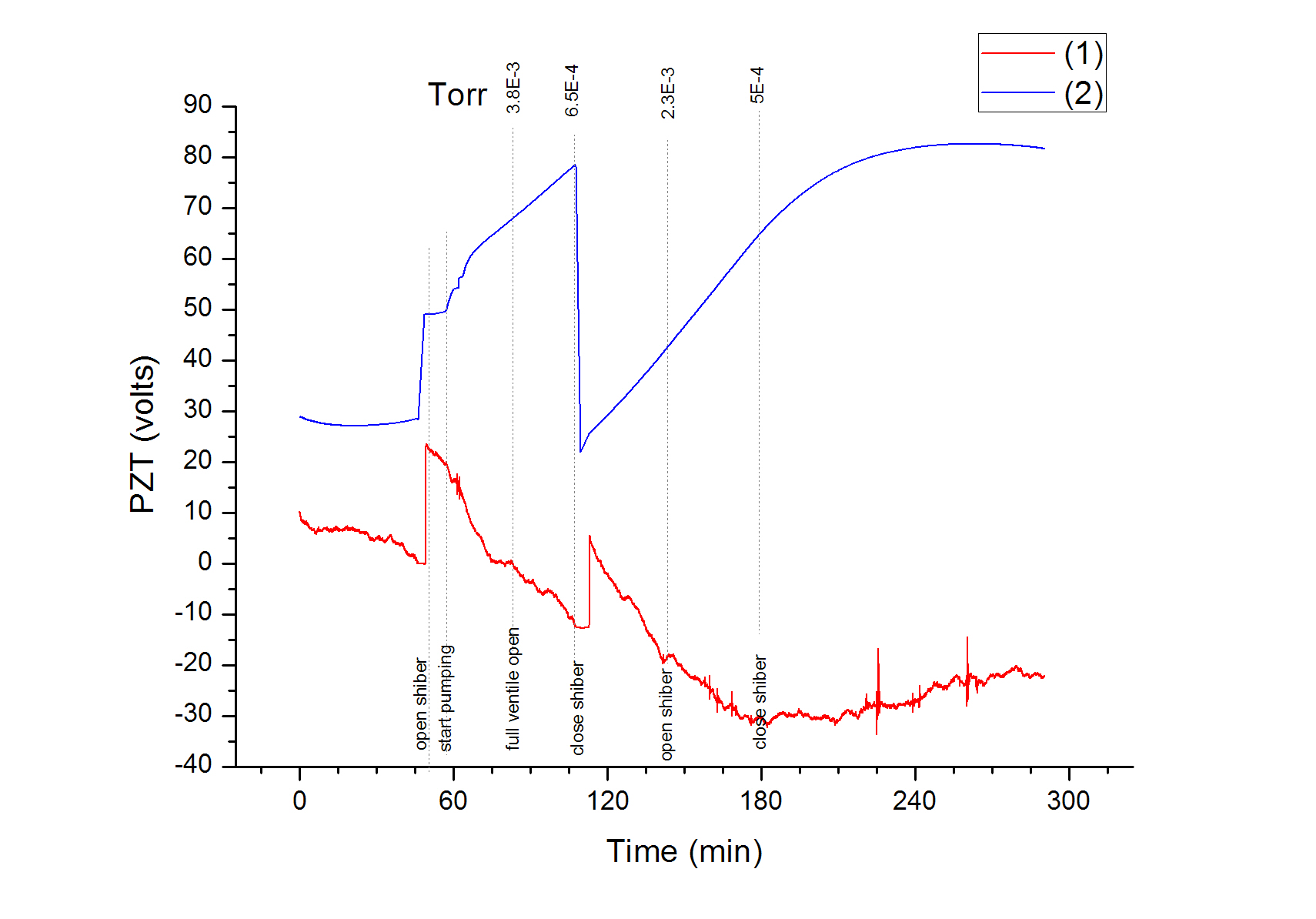}
	\caption{Dependence of the control voltages in both channels of the OGRAN setup on time with a change in the vacuum in the chamber of the reference resonator.)
	}
	\label{fig:pic4}
	\centering
\end{figure}
The data obtained shows that the operating point in each channel was displaced in one direction (in the direction of increasing temperature) regardless of whether the vacuum in the reference resonator chamber improved or deteriorated. This is due to the presence of a person (technicians controlling vacuum) near the setup. Indeed, the process of temperature changes in Fig. 4 began with the appearance of an operator in the OGRAN laboratory at about 30 minutes of recording. The course of the curves turned around together with the departure of the technicians (on the electric train at 220 minutes, its bursts are visible on the red curve). The effect of vacuum is still weakly noticeable in the region of the ``start pumping'' point in the form of a slight irregularity in the curve of the second channel. The jumps of the curves of both channels at the ``open shiber'' and ``close shiber'' points are associated with the breakdown and resetting of the full capture mode due to the mechanical effect on the vacuum chamber when opening and closing the gate valve.

\section[Thermal peak measurements]{Thermal peak measurements}	
The frequency spectrum of OGRAN is shown in Fig. 5a with a spectral resolution of 0.001 Hz. Resonant peak OGRAN is at frequency 1.3 kHz. The peak at a frequency of about 800 Hz with the highest Q corresponds to the bending mode oscillations of the detector. The aluminum bar is suspended horizontally in a steel loop that encompasses the bar in the middle part. The low Q peaks near 2 kHz and 3 kHz correspond to the cutoff frequencies in the feedback loops. The thermal peak with high resolution is shown in Figure 5b. The width of the resonance peak is 0.002 Hz. The signal-to-noise ratio in this measurement is about 30 (with the best setting it reaches 60). To achieve such spectral resolution it is necessary to accumulate a signal for at least 1000 seconds. In this case, three sampling points will fall into the band of the resonance peak in the spectrum.

	\begin{figure}[h!]
	\centering
	\includegraphics[width=1\linewidth]{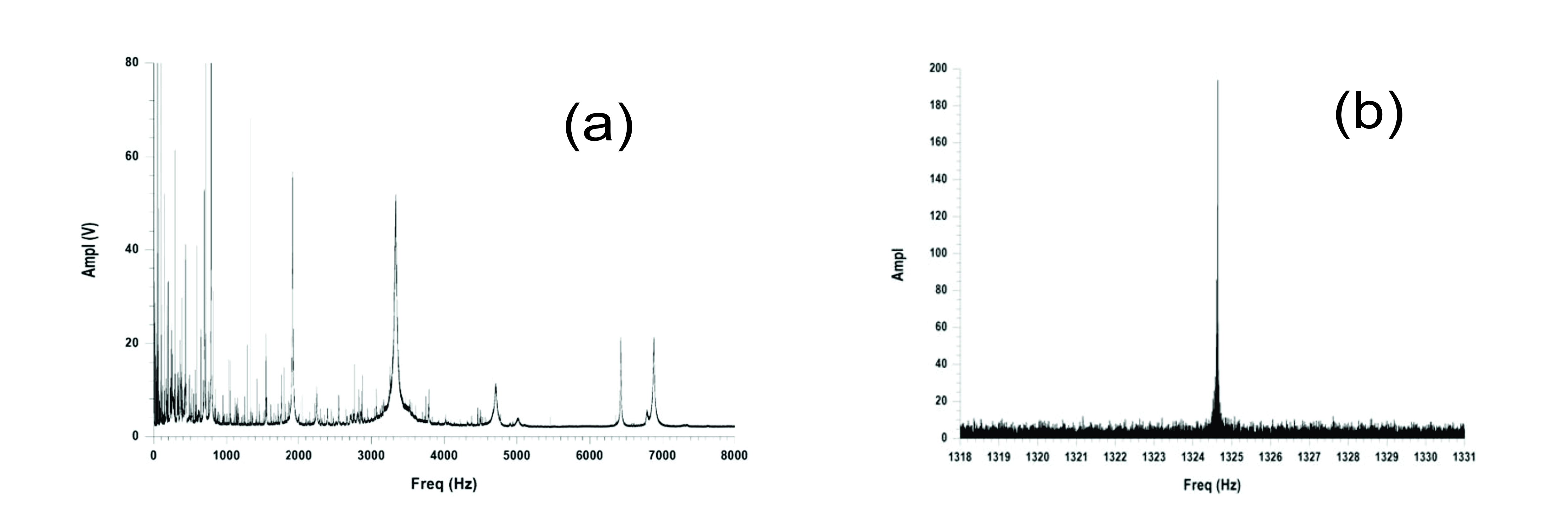}
	\caption{Frequency spectrum of the OGRAN setup (a) and a high-resolution thermal peak (b))
	}
	\label{fig:pic5}
	\centering
\end{figure}
The resonance peak of the OGRAN setup is a reflection of thermal (Brownian) fluctuations of an acoustic detector (bar). The spectral density of this process could be obtained as a result of measurements over a large (ideally infinite) time interval. In practice, each measured realization of a random process is specified only over a finite interval. The spectra of such truncated realizations, called as periodograms, are random reflections of the spectral density \cite{linkk15}. The true value of the spectral density is obtained by adding (averaging) many periodograms (or increasing the measurement time). Fig. 6 shows the frequency spectra in the vicinity of the thermal peak at successive time intervals (1024 s). The total spectrum averaged over the ensemble and normalized to the sample length is highlighted in dark color.

\begin{figure}[h!]
	\centering
	\includegraphics[width=1\linewidth]{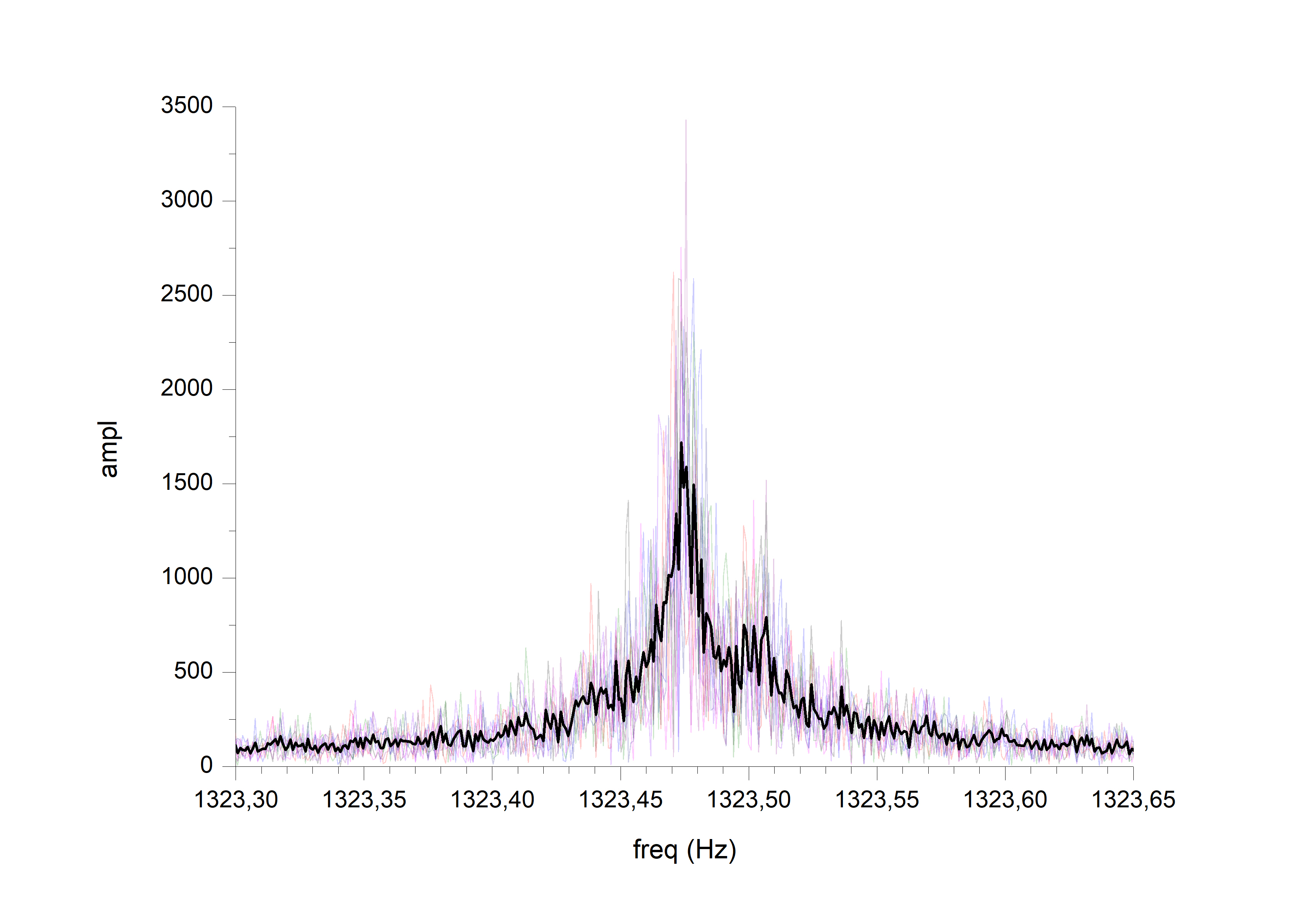}
	\caption{Frequency spectra of the thermal peak measured at separate consecutive 1024 sec intervals with highlighting their average spectrum (bold track), normalized to the sample length
	}
	\label{fig:pic6}
	\centering
\end{figure}

The position of the thermal peak in the stabilized thermal regime changes randomly from sample to sample. Fig.7 shows the results of measuring the center frequency of the thermal peak on successive 1024-second samples.

\begin{figure}[h!]
	\centering
	\includegraphics[width=1\linewidth]{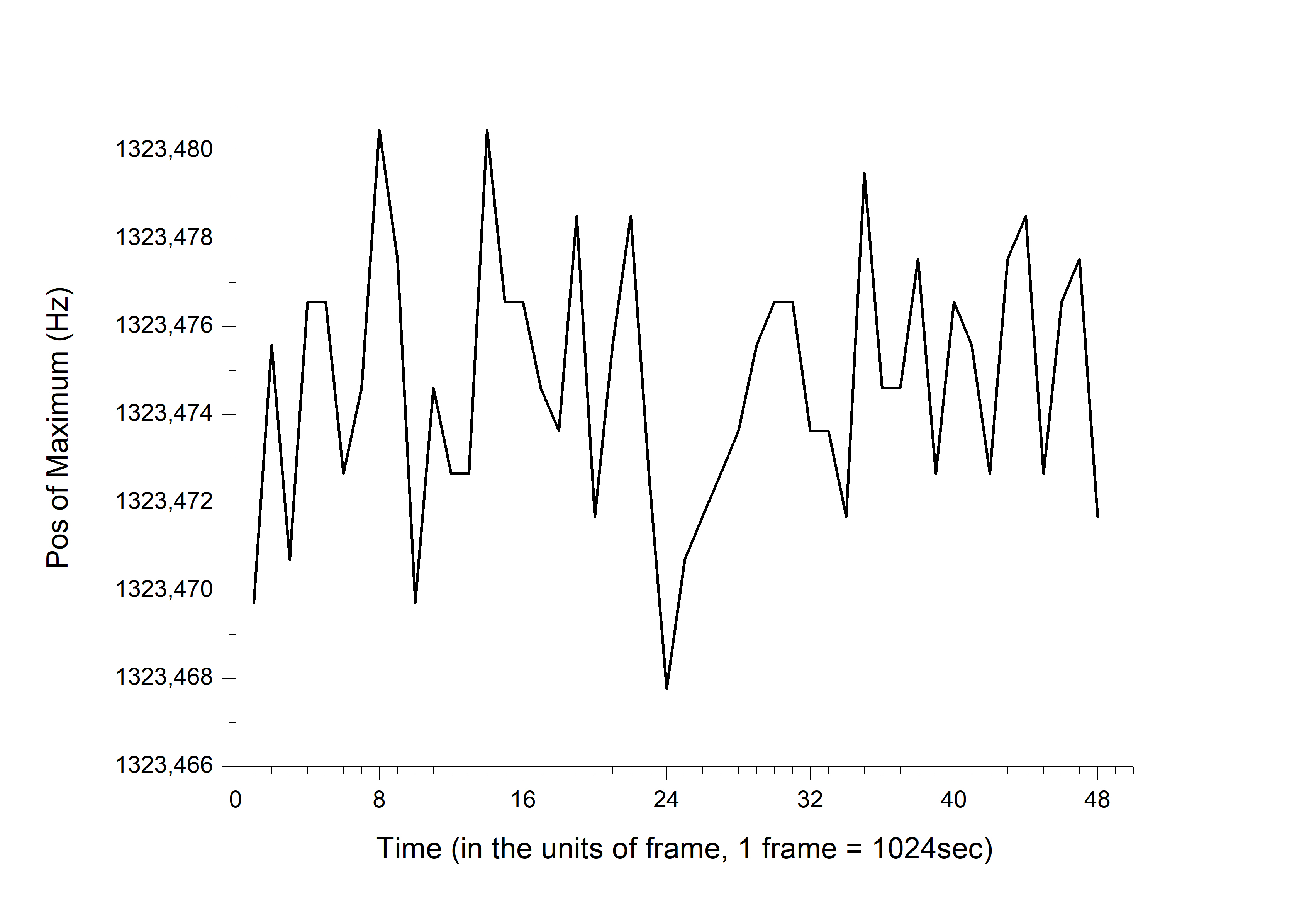}
	\caption{Time shifts of the thermal peak maximum on periodograms in the 14-hour measurement interval. Fourier spectrum time window 1024 sec.
	}
	\label{fig:pic7}
	\centering
\end{figure}

The measurement statistics for the maximum thermal peak are shown in Fig. 8. The histogram fits well with a Gaussian curve.

\begin{figure}[h!]
	\centering
	\includegraphics[width=1\linewidth]{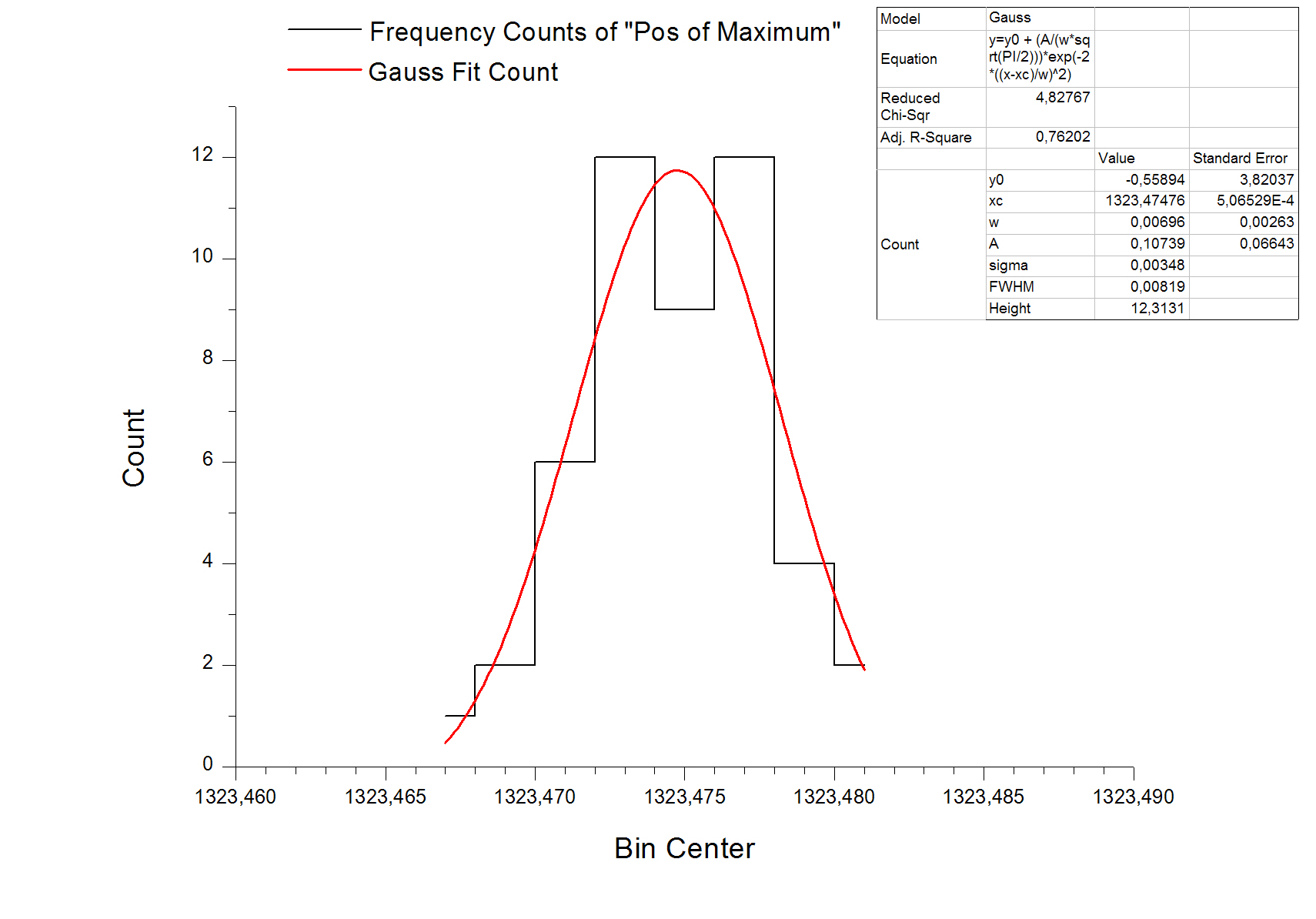}
	\caption{Statistics of displacements of the maximum thermal peak of periodograms in the 14-hour interval of measurements.
	}
	\label{fig:pic8}
	\centering
\end{figure}

\section[Conclusion ]{Conclusion }
The OGRAN setup is launched in the long-term continuous observation mode. The first experiments revealed the importance of the realization of the stable thermal background in the adjacent rooms, as well as the negligible effect of variations in the vacuum in the chambers of detector and discriminator. The time of temperature relaxation inside both FP resonators caused by changing of the environment temperature is about 700 hours. 
Spectral measurements of the thermal peak were carried out. It was shown that for its high-quality reproduction, at least 1000 second signal accumulation is necessary. In this case, several points of the digitized spectrum fall into its narrow bandwidth of 0.002 Hz. Each individual measurement of the spectrum is follow to statistics of measurements of random processes. An increase in the sample length only improves the quality of an individual spectrum, but does not affect the true position of the thermal peak maximum. 

\section[Acknowledgments  ]{Acknowledgments  }
The authors express their gratitude to Academicians S.N. Bagayev and A.M. Cherepashchuk for helpful advice and attention to the work. This work was supported by the RFBR grant 19-29-11010.

\end{document}